\documentstyle[sprocl,epsf]{article}

\begin{document}

\title{Accounting for running $\alpha_s$ for the non-singlet 
components of the structure functions $F_1$ and $g_1$ at small $x$.}

\author{B. I. Ermolaev}
\address{A.F.Ioffe Physico-Technical Institute, 194021
St.Petersburg,Russia}

\author{M. Greco}
\address{Dipartimento di Fisica and INFN, University of Rome III,
Rome, Italy}

\author{S. I. Troyan}
\address{St.Petersburg Institute of
Nuclear Physics, 188300 Gatchina, Russia}

\maketitle\abstracts{Infrared  evolution equations incorporating the
running QCD coupling are constructed and solved for the non-singlet
structure functions $f_{NS}$. Accounting for dropped logs of $x$ in
DGLAP it leads to a scaling-like small $x$ behaviour  $f_{NS} \sim
(\sqrt{Q^2}/x)^a$. In contrast to the leading logarithmic
approximation, intercepts $a$  are numbers and do not contain
$\alpha_s$. It is also shown that the leading logarithmic approximation
may be unreliable for predicting  $Q^2$ -dependence of the DIS
structure functions in the HERA range.}

Non-singlet structure functions, i.e. flavour-dependent contributions to the
deep inelastic structure functions, have been the object
of intensive theoretical
investigation. First, they are interesting because they are
experimentally measurable quantities; second, they are
comparatively technically  simple for analysis, and can be
regarded as a starting ground for a theoretical
description of DIS structure functions.
In the present talk we discuss
the explicit expressions\cite{egt}
for the non-singlet contribution $f^{+}_{NS}$ to
the structure function $F_1$ and for the non-singlet contribution
$f^{-}_{NS}$ to the spin structure function  $g_1$
at $x$. These expressions  account for both leading
(double-logarithmic) and sub-leading (single-logarithmic)
contributions to all orders in QCD coupling and include the
running $\alpha_s$ effects.
Contrary to DGLAP\cite{dglap} and to some other works on
small $x$, we do not use a priori
the standard parametrisation $\alpha_s = \alpha_s(Q^2)$ in our evolution
eqs. Indeed it has been shown recently \cite{egt}
that such a dependence is a good approximation at large $x$
but is not correct when $x$ is small.

As we account for double-logarithmic (DL) and single-logarithmic (SL)
contributions to all orders and
regardless of the arguments, we cannot use the DGLAP eqs.
Instead, we construct and solve two-dimensional infrared evolution equations
(IREE) for  $f_{NS}$ appreciating evolution with respect to $x$ and to $Q^2$.
In the context of this method,  $f^{\pm}_{NS}$ evolves
with respect to the infrared cut-off $\mu$ in the transverse momentum space:
$k_{i \perp}> \mu$ for all virtual particles.
In doing so,  we provide $f^{\pm}_{NS}(x,Q^2)$ with
$\mu$ dependence too. However, it's unavoidable when $\alpha_s$
is running because the standard expression
\begin{equation}
\alpha_s(t) = \frac{1}{b \ln(t/\Lambda_{QCD}^2)}~ ,
\label{alpha}
\end{equation}
is valid only when $t \gg \Lambda_{QCD}^2$ and therefore if we
introduce the infrared cut-off  as
\begin{equation}
\label{mu}
k_{i \perp} >\mu > m_{max} \gg  \Lambda_{QCD}~,
\end{equation}
with $m_{max}$ being the mass of the heaviest involved quark, we can neglect
quark masses and still do not have infrared singularities. Besides
the restrictions imposed by Eq.~(\ref{mu}) $\mu$ is not fixed, so
$f_{NS}$ can evolve  with respect to  $\mu$, eventually
arriving at the following expressions for the non-singlet structure
functions:
\begin{equation}
f^{\pm}_{NS}= \int_{-\imath \infty}^{\imath \infty}
\frac{d \omega}{2\pi \imath}C\left(\frac{1}{x}\right)^{\omega}
\omega \exp\left([(1 + \lambda\omega) F_0^{\pm}] y\right)
\label{fsol}
\end{equation}
where $C$ is an (non-perturbative)
input and $F_0^{\pm}$ are the new anomalous dimensions.
They account for the total resummation of the most essential at small
$x$ NLO contributions of the type $(\alpha_s/\omega^2)^n$ and
$(\alpha_s/\omega)^n$~~$(n = 1,...)$ ,
\begin{equation}
F_0^{\pm} = 2 \left[\omega -\sqrt{\omega^2 -
(1 + \lambda \omega)(A(\omega) + \omega D^{\pm}(\omega))/2\pi^2} \right]
\label{f0}
\end{equation}
where
\begin{equation}
A(\omega) = \frac{4C_F \pi}{b}
\left[ \frac{\eta}{\eta^2 + \pi^2} -  \int_0^{\infty}
\frac{d\rho \exp(-\rho\omega)}{(\rho + \eta)^2 + \pi^2}  \right]~.
\label{a}
\end{equation}
and
\begin{equation}
D^{\pm}(\omega) = \frac{2C_F}{\omega b^2 N }
\int_0^{\infty}
d\rho \exp(-\rho\omega) \ln\Big(\frac{\rho + \eta}{\eta}\Big)
\Big[\frac{\rho + \eta}{(\rho + \eta)^2 + \pi^2} \mp
\frac{1}{\rho + \eta}\big]~.
\label{d}
\end{equation}
We have used in Eqs.~(\ref{a},\ref{d}) the following notations:
$\eta =\ln (\mu^2/\Lambda_{QCD}^2),  ~~\rho = \ln (s/\mu^2), \lambda =
1/2$ and the first coefficient of the $\beta$ -function $b = (11N -
2n_f)/12\pi$.  $A$ corresponds to accounting for running
$\alpha_s$.  $\pi^2$ in denominators appears due to analytical
properties of $\alpha_s(t)$: it must have a non-zero
imaginary part when $t$ is time-like.
$D$ contains the signature-dependent contributions.

Expanding the resummed anomalous dimension $F_0^{\pm}$ into series
in $1/\omega$ we reproduce the singular in $\omega$ terms of
LO and NLO DGLAP- anomalous dimensions where
$\alpha_s(Q^2)C_F/2\pi$ is replaced by $A$.

It is shown in  Refs.~\cite{egt} that $A $ can be approximated by
$\alpha_s(Q^2)C_F/2\pi$ only at large $x$.
Concerning the small- $x$ and large $Q^2$
asymptotics of $f^{\pm}_{NS}$,
Eq.~(\ref{fsol}) reads that
\begin{equation}
\label{as}
f^{\pm}_{NS} \sim x^{-\omega_0^{\pm}}(Q^2/\mu^2)^{\omega_0^{\pm}/2} ,
\end{equation}
with the intercepts
$\omega_0^{\pm}$ being the leading, i.e. the rightmost,
singularities of $F_0^{\pm}$. Eqs.~(\ref{fsol},\ref{f0}) read that
$\omega_0^{\pm}$ are the rightmost roots of
\begin{equation}
\omega^2 -
(1 + \lambda \omega)(A(\omega) + \omega D^{\pm}(\omega))/2\pi^2
= 0~.
\label{omega0}
\end{equation}
Eq.~(\ref{omega0}) contains $n_f, \Lambda_{QCD}$
and $\mu$ as parameters.
Choosing e.g. $n_f~=~3$ and $\Lambda_{QCD} = 0.1$~GeV one can solve
Eq.~(\ref{omega0}) numerically and obtain $\omega_0^{\pm}$ as a function of
$\mu$. The solutions are given in Fig.~1.
Both $\omega_0^{+}$ and $\omega_0^{-}$ acquire imaginary parts at
$\mu < 0.4$~GeV. As besides, for applicability of Eq.~(\ref{alpha})
$\mu$ must be
much greater than $\Lambda_{QCD}$, we think that the region $\mu <
0.4$~GeV is beyond control of our approach. Both  $\omega_0^{+}$ and
$\omega_0^{-}$ are maximal at $\mu\approx 1$~Gev and slowly decrease
with $\mu$ increasing.  Therefore we can estimate values of the
intercepts as
\begin{equation}
\Omega_0^{+} = 0.37,~~~~ \Omega_0^{-} = 0.4 ~.
\label{Omega}
\end{equation}

It is interesting that
this result was independently confirmed\cite{kat} recently by
extrapolating of fits for $f_3$ into small $x$ region.
Eq.~(\ref{Omega}) was obtained from  Eq.~(\ref{omega0}) which contains
$\pi^2$ -terms. Basically, they are beyond of control of logarithmic accuracy
and might be dropped. With $\pi^2$ -terms dropped, we obtain
the smooth curves for  $\omega_0^{\pm}$ depicted in Fig.~1.
These curves show that $\pi^2$ -terms can be easily neglected for the
values of $\mu$ greater than $\mu_0 = 5.5$ GeV. However, $\mu_0^2 = 30$
Gev$^2$  corresponds to the HERA $Q^2$ range. Then
Eq.~(\ref{Omega}) immediately implies that, with such a big $\mu_0$,
the logarithmic accuracy is not enough to obtain a correct $Q^2$
dependence in the HERA range. On the other hand, it also explains why DLA
estimates
$\alpha_s = \alpha_s(Q^2)$ may be correct for predicting the $x$
dependence: indeed, in DLA where the coupling is fixed,
one should use rather $\alpha_s = \alpha_s(\mu_0^2)$ than
$\alpha_s = \alpha_s(Q^2)$ as taken from DGLAP,
but as it happens that $\mu_0^2 = Q^2$ in the HERA range, both estimates
coincide.
\begin{figure}[t]
\begin{center}
\begin{picture}(320,220)
\put(0,20){
\epsfbox{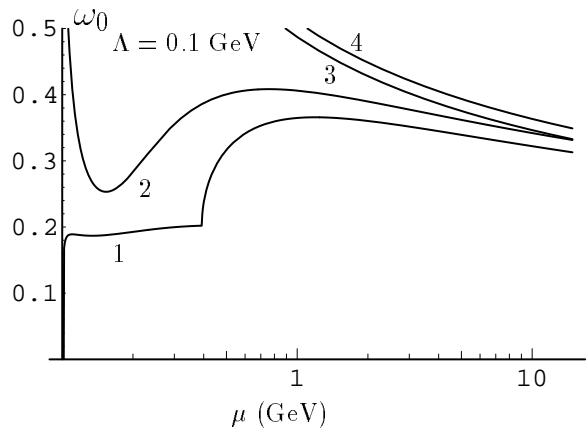}
}
\end{picture}
\end{center}
\caption{Dependence of the intercept $\omega_0$ on infrared cutoff
$\mu$ at $\Lambda_{QCD}=0.1$~GeV: 1-- for $f_1^{NS}$; 2-- for
$g_1^{NS}$; 3-- and 4-- for $f_1^{NS}$ and $g_1^{NS}$ respectively
without account of $\pi^2$-terms.
}
\end{figure}

\section*{Acknowledgments}
This work supported in part by grants INTAS-97-30494, RFBR 00-15-96610 and
the EU QCDNET contract FMRX-CT98-0194.

\end{document}